\documentclass{article}
\usepackage{amssymb,amsmath,epsfig,epstopdf,color,graphicx,subfigure}
\usepackage{epstopdf,array}
\newcolumntype{L}{>{\centering\arraybackslash}m{3cm}}

\usepackage{arxiv}

\usepackage[utf8]{inputenc} % allow utf-8 input
\usepackage[T1]{fontenc}    % use 8-bit T1 fonts
\usepackage{hyperref}       % hyperlinks
\usepackage{url}            % simple URL typesetting
\usepackage{booktabs}       % professional-quality tables
\usepackage{amsfonts}       % blackboard math symbols
\usepackage{nicefrac}       % compact symbols for 1/2, etc.
\usepackage{microtype}      % microtypography

\usepackage{amsmath,amssymb,amsfonts}
\usepackage{algorithmic}
\usepackage{graphicx}
\usepackage{textcomp}
\usepackage{comment,color}
\usepackage{enumitem}
\usepackage{epsfig}
\usepackage{caption}
\usepackage{verbatim,bm}
\usepackage{epstopdf,array,mathtools,url,slashbox,subfigure}

\newcolumntype{L}{>{\centering\arraybackslash}m{3cm}}
\DeclarePairedDelimiter\norm{\lVert}{\rVert}%
\makeatletter
\let\oldabs\abs
\def\abs{\@ifstar{\oldabs}{\oldabs*}}
\let\oldnorm\norm
\def\norm{\@ifstar{\oldnorm}{\oldnorm*}}
\makeatother
\title{PINNPStomo: Simultaneous P- and S-wave seismic traveltime tomography using physics-informed neural networks with a new factored eikonal equation}

\author{
  Chao Song\\
 Department of Geophysics, College of Geo-\\
 Exploration Science and Technology, \\
 Jilin University
    \And
   Hang Geng\\
 Department of Geophysics, College of Geo-\\
 Exploration Science and Technology, \\
 Jilin University
    \And 
  Umair bin Waheed\\
  Department of Geosciences\\
  King Fahd University of Petroleum and Minerals
    \And  
  Cai Liu\\
 Department of Geophysics, College of Geo-\\
 Exploration Science and Technology, \\
 Jilin University
    \And  
}

\begin{document}
\maketitle

\begin{abstract}

Seismic tomography has long been an effective tool for constructing reliable subsurface structures. However, simultaneous inversion of P- and S-wave velocities presents a significant challenge for conventional seismic tomography methods, which depend on numerical algorithms to calculate traveltimes. A physics informed neural network (PINN)-based seismic tomography method (PINNtomo) has been proposed to solve the eikonal equation and construct the velocity model. Leveraging the powerful approximation capabilities of neural networks, we propose extending PINNtomo to perform multiparameter inversion of P- and S-wave velocities jointly, which we refer to as PINNPStomo. In PINNPStomo, we employ two neural networks: one for the P- and S-wave traveltimes, and another for the P- and S-wave velocities. By optimizing the misfits of P- and S-wave first-arrival traveltimes calculated from the eikonal equations, we can obtain the predicted P- and S-wave velocities that determine these traveltimes. Recognizing that the original PINNtomo utilizes a multiplicative factored eikonal equation, which depends on background traveltimes corresponding to a homogeneous velocity at the source location. We propose a new factored eikonal equation for PINNPStomo to eliminate this dependency. The proposed PINNPStomo, incorporating the new factored eikonal equation, demonstrates superior convergence speed and multiparameter inversion accuracy. We validate these improvements using 2D Marmousi and 2D/3D Overthrust elastic velocity models across three different seismic data acquisition geometries.

\end{abstract}

% keywords can be removed

\section{Introduction}

Seismic tomography is an essential technique for subsurface parameter building across various scales. In global or regional seismology, seismic tomography enables imaging of the Earth's inner structure \cite{van1997evidence,rawlinson2010seismic}. In exploration seismology, it provides a reliable background velocity model for subsequent high-resolution seismic imaging processes, such as reverse time migration (RTM) and full-waveform inversion (FWI) \cite{levin1984principle,tarantola1984inversion,virieux2009overview,biondi2013tomographic,alkhalifah2014tomography,yao2019extraction}. The high-frequency approximation ray theory imaging method \cite{zhao1992tomographic}, the finite-frequency imaging method \cite{dahlen2000frechet}, and the traveltime tomography method based on the eikonal equation \cite{rawlinson2004wave} are three of the most widely used representative first-arrival traveltime tomography methods. Compared to the above two kinds of ray-theory-based tomography methods, the equation based tomography method does not require ray tracing to solve the first-arrival traveltimes in complex models \cite{taillandier2009first,waheed2016first}. By optimizing the residuals between simulated and observed traveltimes, it can directly obtain the velocity gradient, thereby improving the stability of velocity inversion \cite{leung2006adjoint}.

Solving the eikonal equation can simulate the first-arrival traveltimes of seismic waves used to achieve seismic tomography \cite{zhou2023topography}. The eikonal equation is a non-linear first-order partial differential equation (PDE). The fast marching method (FMM) and fast sweeping method (FSM) are primarily used to solve the eikonal equation \cite{sethian1996fast,zhao2005fast}. Both methods are fundamentally based on the finite difference scheme, and their accuracy depends on the model meshing schemes. Additionally, the eikonal equation is limited to simulating the traveltime of one-component (P or S) body waves.

Elastic subsurface parameter inversion provides key insights into understanding the Earth's structure and determining oil and gas reservoirs. Elastic FWI is gaining popularity due to its improved accuracy in interpreting shear and converted waves in seismic data \cite{brossier2009seismic,zhang2019normalized}. Simultaneous P- and S-wave seismic tomography would be of vital benefit in obtaining elastic background models. In practical applications, explosive sources can excite both P-waves and S-waves. Borehole seismic data are less affected by surface wave interference and have a high signal-to-noise ratio, enabling the picking of first-arrival times of P-waves and S-waves and completing P-wave and S-wave travel-time tomography. However, traditional eikonal equation tomography methods based on numerical algorithms can only be applied to one-component P- or S-waves. Thus, these numerical algorithms cannot quickly simulate P- and S-wave traveltimes and invert velocity models simultaneously, limiting the efficiency of elastic velocity inversion.

In recent years, the rapid development of deep learning technology has made it an effective tool in geophysics, with widespread applications \cite{mousavi2022deep,wu2023sensing}. Physics-Informed Neural Networks (PINNs) are a novel framework for solving partial differential equations based on deep learning theory \cite{raissi2019physics,karniadakis2021physics}. Compared to traditional numerical algorithms, PINN is a mesh-free method for solving partial differential equations. By using coordinate values as inputs and physical equations as the loss function, the training process can directly yield the target function. In seismic exploration, PINNs have been applied to time-domain and frequency-domain seismic wavefield simulation and inversion in various media, demonstrating convincing effectiveness and strong flexibility \cite{moseley2020solving,alkhalifah2021wavefield,song2021solving,huang2022pinnup,rasht2022physics,song2022wavefield,
song2023simulating,wu2023helmholtz,chai2024modeling}. In seismic traveltime simulation, PINNs have shown superior accuracy in solving the eikonal equation compared to numerical solvers \cite{smith2020eikonet,bin2021pinneik}. Building on this foundation, \cite{waheed2021pinntomo} developed a PINN-based traveltime tomography method, PINNtomo, to efficiently build the background P-wave velocity. Researchers have proposed using Bayesian theory, data hard-constraint, and well-log velocity constraint to further improve the accuracy of PINNtomo \cite{agata2023bayesian,gou2023bayesian,taufik2023robust,zhao2024smoothness}.
 
Leveraging the flexibility of PINNs, we develop a PINN-based simultaneous P- and S-wave traveltime tomography method, namely PINNPStomo. We use two independent neural networks to represent the P- and S-wave traveltimes and velocities, respectively. The loss function combines the physics loss corresponding to the eikonal equation and the data loss from the traveltime discrepancy for both P- and S-waves. With a limited number of training epochs, we can achieve training convergence and simultaneously estimate the P- and S-wave velocities. The original PINNtomo method used the multiplicative factored eikonal equation as the physics loss to avoid the source singularity \cite{waheed2021pinntomo}. This method requires prior information about the velocity values at the source locations to calculate the background traveltimes. However, obtaining this prior information is challenging, especially when the sources are distributed in the subsurface. To resolve this issue, we propose a new factored eikonal equation for PINNPStomo, which is independent of the background traveltimes. We find that PINNPStomo with this new factored eikonal equation can estimate P- and S-wave velocities accurately and efficiently. We demonstrate these features on 2D and 3D velocity models with three different seismic acquisition geometry setups.

\section{Theory}

\subsection{The eikonal equation and its multiplicative factored form}

Seismic P- and S-wave first-arrival traveltimes can be simulated by solving a first-order, hyperbolic form of partial differential equation (PDE), namely eikonal equation given by \cite{julian1977three,cerveny2001seismic}:

\begin{eqnarray}
\left | \nabla T_{p,s}(\mathbf{x}) \right |^{2}=\frac{1}{v_{p,s}^{2}(\mathbf{x})},\; \; \forall \mathbf{x}\in \Omega ,
\label{eqn:eq1}
\end{eqnarray}

where $T_{p,s}(\mathbf{x})$ represents the traveltimes of P or S waves in the domain of $\Omega$, and $v_{p,s}(\mathbf{x})$ represents the P- or S-wave velocity model that determines $T_{p}(\mathbf{x})$ and $T_{s}(\mathbf{x})$. $\mathbf{x}=\left \{ x, y ,z\right \}$ specifies the Euclidean coordinates for spatial positions. $\nabla=\frac{\partial}{\partial x}+\frac{\partial}{\partial y}+\frac{\partial}{\partial z}$ denotes the operator of the first-order spatial derivative. In the source location $\mathbf{x_{s}}=\left \{ x_{s}, y_{s},z_{s}\right \}$, the seismic first-arrival traveltimes for both P- ans S-wave should be zero, given by: $T_{p,s}(\mathbf{x_{s}})=0$. This point-source singularity issue will cause inaccurate traveltime solutions around the source location. To mitigate this issue, the factored form of the eikonal equation can be used by decomposing the $T_{p,s}(\mathbf{x})$ as \cite{fomel2009fast}: 
\begin{eqnarray}
T_{p,s}(\mathbf{x})=T_{p0,s0}(\mathbf{x})\tau_{p,s}(\mathbf{x}),
\label{eqn:eq2}
\end{eqnarray}
where $T_{p0,s0}(\mathbf{x})$ denotes the P- ($T_{p0}$) or S-wave ($T_{s0}$) background traveltime term from a constant P- $(v_{p0})$ or S-wave velocity $(v_{s0})$ value. $T_{p0,s0}(\mathbf{x})$ can be easily calculated by:
\begin{eqnarray}
T_{p0,s0}(\mathbf{x})=\frac{\left | \mathbf{x}-\mathbf{x_{s}} \right |}{v_{p0,s0}},
\label{eqn:eq3}
\end{eqnarray}
where $\left | \mathbf{x}-\mathbf{x_{s}} \right |$ denotes the seismic wave travelling distance from the point $\mathbf{x}$ to the source point $\mathbf{x_{s}}$. The background velocities $v_{p0}$ and $v_{p0}$ are sampled at the source locations. For every different source, $v_{p0,s0}$ is usually different. $\tau_{p,s}$ denotes the traveltime factor for P or S wave, which is unitless. Mathematically, the traveltime factor $\tau$ (for either P- or S-wave) is equal to:
\begin{eqnarray}
\tau=\frac{T}{T_{0}}=\frac{\left | \mathbf{x}-\mathbf{x_{s}} \right |/v^{'}}{\left | \mathbf{x}-\mathbf{x_{s}} \right |/v_{0}}=\frac{v_{0}}{v^{'}},
\label{eqn:eq4}
\end{eqnarray}
where $v^{'}$ can be considered as the average velocity along the wavepath between point $\mathbf{x}$ to the source location $\mathbf{x_{s}}$. As the value of $v^{'}$ naturally falls within the range of $\left [ v_{min}, v_{max} \right ]$, where $v_{max}$ and $v_{min}$ denote the maximum and minimum velocities of the velocity model (for either P- or S-wave). The range of $\tau$ results in between $\frac{v_{0}}{v_{max}}$ and $\frac{v_{0}}{v_{min}}$. For each source, the range of $\tau$ varies with the change of velocity value at the source $v_{0}$. 

By incorporating the factored form of traveltime $T_{p0,s0}(\mathbf{x})\tau_{p,s}(\mathbf{x})$ in equation \ref{eqn:eq1}, the factored form of the eikonal equation can be expressed as:
\begin{eqnarray}
&& T_{p0,s0}^{2}(\nabla\tau_{p,s})^{2}+\tau_{p,s}^{2}(\nabla T_{p0,s0})^{2}+2\tau_{p,s}T_{p0,s0} (\nabla T_{p0,s0} \nabla \tau_{p,s})=\frac{1}{v_{p,s}^{2}},\nonumber\\
&& \tau_{p,s}(\mathbf{x_{s}})=1.
\label{eqn:eq5}
\end{eqnarray}

\subsection{Physics-informed neural network}

\begin{figure}
\begin{center}
\includegraphics[width=1.0\textwidth]{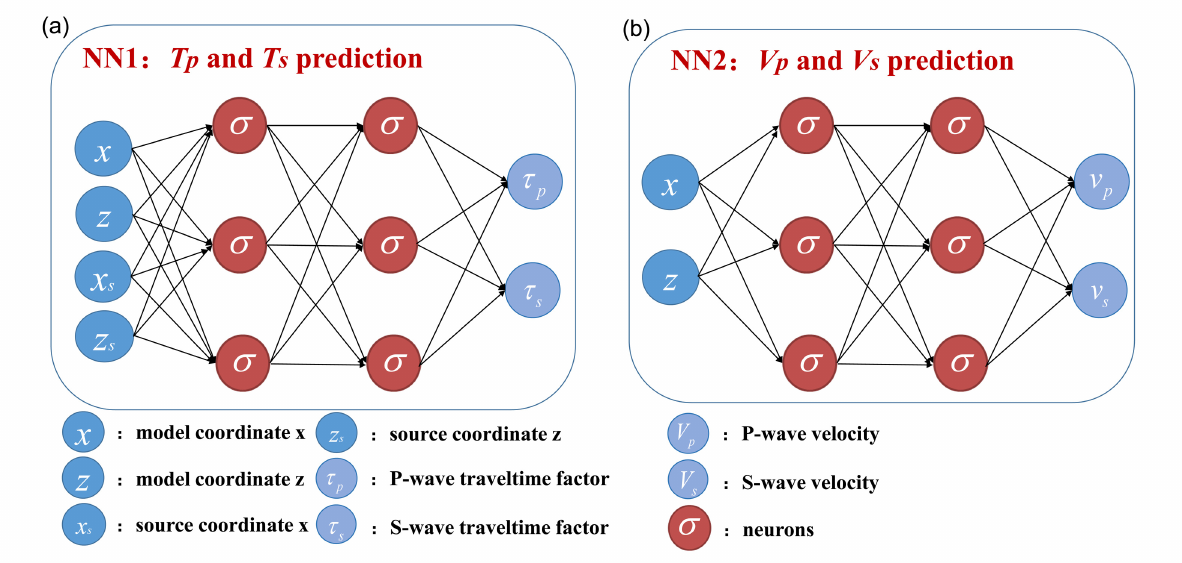} 
\caption{(a) NN1 for P- and S-wave traveltime prediction; (b) NN2 for P- and S-wave velocity inversion.}
\label{fig:PINN_network}
\end{center}   
\end{figure}

In seismic traveltime tomography, we estimate the velocity model by minimizing the traveltime misfit between the observed and predicted seismic data. According to the universal approximation theorem of neural networks (NNs) \cite{hornik1991approximation,leshno1993multilayer}, we use two separate NNs to approximate P-, S-wave traveltime factors and velocity models. As illustrated by Fig. \ref{fig:PINN_network}, $\mathit{NN}1$ is built for multi-source P- and S-wave traveltime prediction, mathematically expressed as: $\mathit{NN}1(\theta_{\tau})\approx \left [ \widetilde{\tau_{p}},\widetilde{\tau_{s}} \right ]$. $\left [ \widetilde{\tau_{p}},\widetilde{\tau_{s}} \right ]$ are outputs of $\mathit{NN}1$. $\theta_{\tau}$ is trainable parameters of $\mathit{NN}1$. The inputs of NN1 include the coordinates of the model, as well as, the source locations. While  $\mathit{NN}2$ is built for predicting the P- and S-wave velocities $\left [ \widetilde{v_{p}},\widetilde{v_{s}} \right ]$, given by $\mathit{NN}2(\theta_{v})\approx \left [ \widetilde{v_{p}},\widetilde{v_{s}} \right ]$. $\theta_{v}$ is the trainable parameters of $\mathit{NN}2$. The inputs of $\mathit{NN}2$ only contain the model coordinates as they are source-independent. 

To predict P- and S-wave velocity models simultaneously, we incorporate the outputs of $\mathit{NN}1$ and $\mathit{NN}2$ to form one loss function expressed as:

\begin{eqnarray}
&& \jmath(\theta_{\tau_{p,s}},\theta_{v_{p,s}}) =\frac{1}{N}\sum_{i=1}^{N}\left \| L_{p}(\widetilde{\gamma}_{p},\widetilde{v}_{p}) \right \|_{2}^{2}+ \frac{1}{N_{s}N_{r}}\sum_{is=1}^{N_{s}}\sum_{ir=1}^{N_{r}}\left \| T_{p0}^{(is)}(\mathbf{x_{r}})\widetilde{\tau_{p}}^{(is)}(\mathbf{x_{r}}) -T_{p}^{(is,ir)}\right \|_{2}^{2} + \nonumber\\
&& \frac{1}{N}\sum_{i=1}^{N}\left \| L_{s}(\widetilde{\gamma}_{s},\widetilde{v}_{s}) \right \|_{2}^{2} +\frac{1}{N_{s}N_{r}}\sum_{is=1}^{N_{s}}\sum_{ir=1}^{N_{r}}\left \| T_{s0}^{(is)}(\mathbf{x_{r}})\widetilde{\tau_{s}}^{(is)}(\mathbf{x_{r}}) -T_{s}^{(is,ir)}\right \|_{2}^{2},
\label{eqn:eq6}
\end{eqnarray}
with
\begin{eqnarray}
&& L_{p}(\widetilde{\gamma}_{p},\widetilde{v}_{p}):=T_{p0}^{2}(\nabla\widetilde{\tau_{p}})^{2}+\widetilde{\tau_{p}}^{2}(\nabla T_{p0})^{2}+2\widetilde{\tau_{p}}T_{p0} (\nabla T_{p0} \nabla \widetilde{\tau_{p}})-\frac{1}{\widetilde{v_{p}}^{2}},\nonumber\\
&& L_{s}(\widetilde{\gamma}_{s},\widetilde{v}_{s}):=T_{s0}^{2}(\nabla\widetilde{\tau_{s}})^{2}+\widetilde{\tau_{s}}^{2}(\nabla T_{s0})^{2}+2\widetilde{\tau_{s}}T_{s0} (\nabla T_{s0} \nabla \widetilde{\tau_{s}})-\frac{1}{\widetilde{v_{s}}^{2}},
\label{eqn:eq7}
\end{eqnarray}

where $N$ is the number of the input coordinates randomly selected within the model domain. In equation \ref{eqn:eq6}, the observed first-arrival traveltimes of P-($T_{p}^{is,ir}$) and S waves ($T_{s}^{is,ir}$) from the $is-$th source and $ir-$th receiver are used. $N_{s}$ and $N_{r}$ are numbers of sources and receivers, and $\mathbf{x_{r}}$ denotes the coordinates of the receivers. By optimizing the loss function in equation 5, wwe can obtain P- and S-wave traveltimeIts range should bes that fit the observed data and satisfy the eikonal equations corresponding to the predicted P- and S-wave velocity models. The prediction of the traveltimes $T$ rely on the background traveltimes $T_{0}$, which implies that the velocity value at the source location must be provided. However, when sources are distributed in the subsurface, the exact velocity values at the source locations are often unknown, which can pose challenges for PINNPStomo.

\subsection{A new factored form of eikonal equation for PINNPStomo}

To mitigate the dependency of velocity values at the source locations, we develop a new factored form of eikonal equation, specifically for PINNPStomo. We decompose the traveltime $T_{p,s}(\mathbf{x})$ as $T_{p,s}(\mathbf{x})=\gamma_{p,s} \left | \mathbf{x}-\mathbf{x_{s}} \right |$. In this case, $\gamma_{p,s}=\frac{T_{p,s}(\mathbf{x})}{\left | \mathbf{x}-\mathbf{x_{s}} \right |}$, which is independent of $v_{p0,s0}$. 
The new factored eikonal equation is expressed as:
\begin{eqnarray}
R^{2}(\nabla \gamma_{p,s})^{2}+\gamma_{p,s}^{2}(\nabla R)^{2}+2\gamma_{p,s} R (\nabla R \nabla \gamma_{p,s})=\frac{1}{v_{p,s}^{2}},
\label{eqn:eq8}
\end{eqnarray}
where $R=\left | \mathbf{x}-\mathbf{x_{s}} \right |$. For simultaneous P- and S-wave PINNtomo using this new factored eikonal equation, we use the same $\mathit{NN}2(\theta_{v})$ to predict the P- and S-wave velocity models $\left [ \widetilde{v_{p}},\widetilde{v_{s}} \right ]$. The physics loss term of the new PINNPStomo uses the new factored eikonal equation in equation \ref{eqn:eq8}. We use $R^{(is)}(\mathbf{x_{r}})\widetilde{\gamma}_{p,s}$ to obtain the predicted P- and S-wave traveltimes for the $is-$th source at the receiver locations $\mathbf{x_{r}}$. Considering the physics loss and the traveltime loss, the loss function of PINNPStomo using the new factored eikonal equation is given by:
 
\begin{eqnarray}
&& \jmath(\theta_{\gamma_{p,s}},\theta_{v_{p,s}}) =\frac{1}{N}\sum_{i=1}^{N}\left \| L_{p}(\widetilde{\gamma}_{p},\widetilde{v}_{p}) \right \|_{2}^{2} + \frac{1}{N_{s}N_{r}}\sum_{is=1}^{N_{s}}\sum_{ir=1}^{N_{r}}\left \| R^{(is)}(\mathbf{x_{r}})\widetilde{\gamma_{p}}^{(is)}(\mathbf{x_{r}}) -T_{p}^{(is,ir)}\right \|_{2}^{2}+\nonumber\\
&&  \frac{1}{N}\sum_{i=1}^{N}\left \| L_{s}(\widetilde{\gamma}_{s},\widetilde{v}_{s}) \right \|_{2}^{2} + \frac{1}{N_{s}N_{r}}\sum_{is=1}^{N_{s}}\sum_{ir=1}^{N_{r}}\left \| R^{(is)}(\mathbf{x_{r}})\widetilde{\gamma_{s}}^{(is)}(\mathbf{x_{r}}) -T_{s}^{(is,ir)}\right \|_{2}^{2},
\label{eqn:eq9}
\end{eqnarray}
with
\begin{eqnarray}
&& L_{p}(\widetilde{\gamma}_{p},\widetilde{v}_{p}):=R^{2}(\nabla\widetilde{\gamma_{p}})^{2}+\widetilde{\gamma_{p}}^{2}(\nabla R)^{2}+2\widetilde{\gamma_{p}}R (\nabla R \nabla \widetilde{\gamma_{p}})-\frac{1}{\widetilde{v_{p}}^{2}},\nonumber\\
&& L_{s}(\widetilde{\gamma}_{s},\widetilde{v}_{s}):=R^{2}(\nabla\widetilde{\gamma_{s}})^{2}+\widetilde{\gamma_{s}}^{2}(\nabla R)^{2}+2\widetilde{\gamma_{s}}R (\nabla R \nabla \widetilde{\gamma_{s}})-\frac{1}{\widetilde{v_{s}}^{2}}.
\label{eqn:eq10}
\end{eqnarray}

Theoretically, $\gamma_{p,s}$ can be interpreted as the average P- or S-wave slowness along the wavepath between the point $\mathbf{x}$ to the source location $\mathbf{x_{s}}$ with unit of $s/km$. Its range should be $\left [ \frac{1}{v_{max}}, \frac{1}{v_{min}} \right ]$, which applies for all sources. Unlike $\tau$, which varies for each source, the range of $\gamma$ is more consistent and easier to retrieve for neural networks. This feature provides an additional advantage to the proposed PINNPStomo with the new eikonal equation: it will accelerate training convergence.

\subsection{Training setup}

We train the networks using the Adam optimizer with a mini-batch scheme. For all the experiments implemented in this paper, $\mathit{NN}1$ contains 8 hidden layers and 64 neurons in each layer.  While $\mathit{NN}2$ has the same number of hidden layers as $\mathit{NN}1$, but half the number of neurons in each layer, as$\mathit{NN}2$ is source-independent with simpler outputs. The exponential linear unit (ELU) function is used to activate the neurons within the hidden layers. In the last layer, which connects the last hidden layer to the output, we employ the sigmoid activation function to ensure that the output range is within $[0, 1]$. PINNPStomo does not require any information about the initial P- and S-wave velocity models. The initial velocity models used in all the examples in this paper are randomly initialized by $\mathit{NN}2$.

\section{Results}

In seismic land surveys, identifying the first-arrival traveltimes of S-waves is challenging due to the strong energy of surface waves. We evaluate the proposed PINNPStomo method on three types of seismic data where S-wave first arrivals can be identified: walkaway vertical seismic profile (VSP) data, crosswell data, and passive seismic data. We generate synthetic seismic data of these types on 2D Marmousi and 2D/3D Overthrust models to validate the proposed method.

\subsection{Walkaway VSP data for the Marmousi model}

\begin{figure}
\begin{center}
\includegraphics[width=1.0\textwidth]{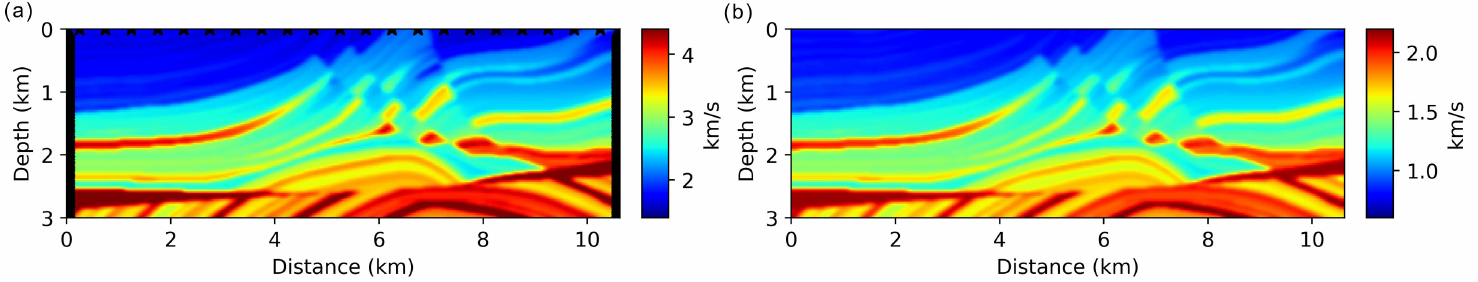} 
\caption{True (a) P- and (b) S-wave elastic Marmousi velocity models. The sources are placed on the surface indicated by the $\star$ sign; the receivers are distributed on both edges of the model with a 25-m interval.}
\label{fig:mar_vp_vs_tomo_vsp_true}
\end{center}   
\end{figure}

We first evaluate the effectiveness of the proposed PINNPStomo framework for inverting P- and S-wave velocity models using the elastic Marmousi model. The size of the model is $426 \times 121$, with a spatial sampling interval of 25 m in both vertical and horizontal directions. The P- and S-wave velocity models are shown in Figs. \ref{fig:mar_vp_vs_tomo_vsp_true}a and \ref{fig:mar_vp_vs_tomo_vsp_true}b, respectively. To simulate the walkaway VSP seismic data acquisition geometry, we place 21 sources evenly distributed on the surface to generate observed P- and S-wave first-arrival traveltimes using FMM. Additionally, we position two wells on the edges of the model, each containing 121 receivers.

\begin{figure}
\begin{center}
\includegraphics[width=0.75\textwidth]{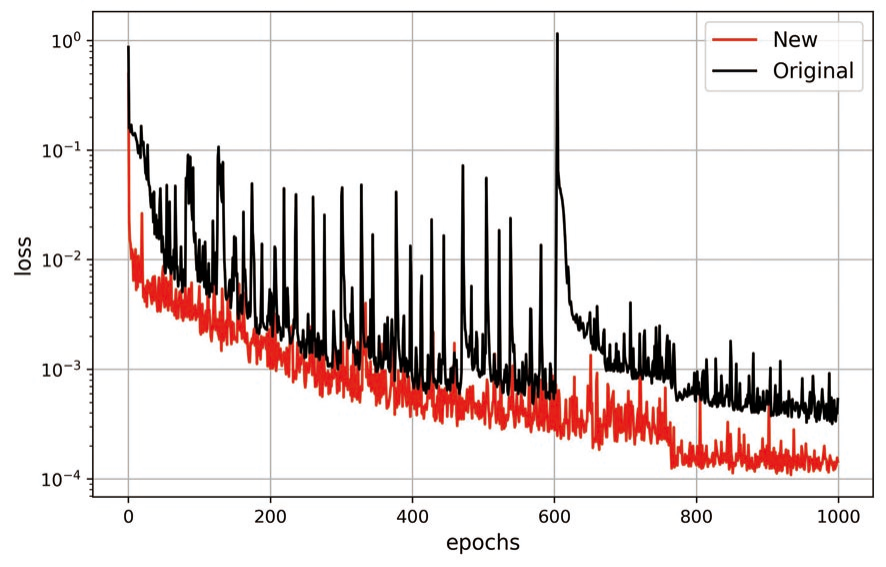} 
\caption{Training loss curves for the new factored eikonal equation (red line) and the original one with (blue line) and without (black line) $v_{p0,s0}$ perturbation.}
\label{fig:mar_loss_comp_vsp_random_noi}
\end{center}   
\end{figure}

In this example, we generate 200,000 sets of $\left \{ x,z,x_{s} \right \}$ consisting of random values within the model domain's range as input coordinates for $\mathit{NN}1$, and we select the corresponding $\left \{ x, z \right \}$ as inputs of $\mathit{NN}2$. With the same hyperparameters, we train the networks with 1000 epochs for the original and new PINNPStomo. Assuming the P- and S-wave velocity values $v_{p0,s0}$ at the source locations are given, we use the original factored eikonal (equation \ref{eqn:eq5}) as the physics constraint to train the networks. Then, we use the proposed new factored eikonal (equation \ref{eqn:eq8}), which does not require the information of $v_{p0,s0}$, as the physics constraint in the loss function. The training loss curves of three different strategies are shown in Fig. \ref{fig:mar_loss_comp_vsp_random_noi}. Obviously, the training loss of the new PINNPStomo strategy decreases faster and reaches the lowest loss. 

\begin{figure}
\begin{center}
\includegraphics[width=1.0\textwidth]{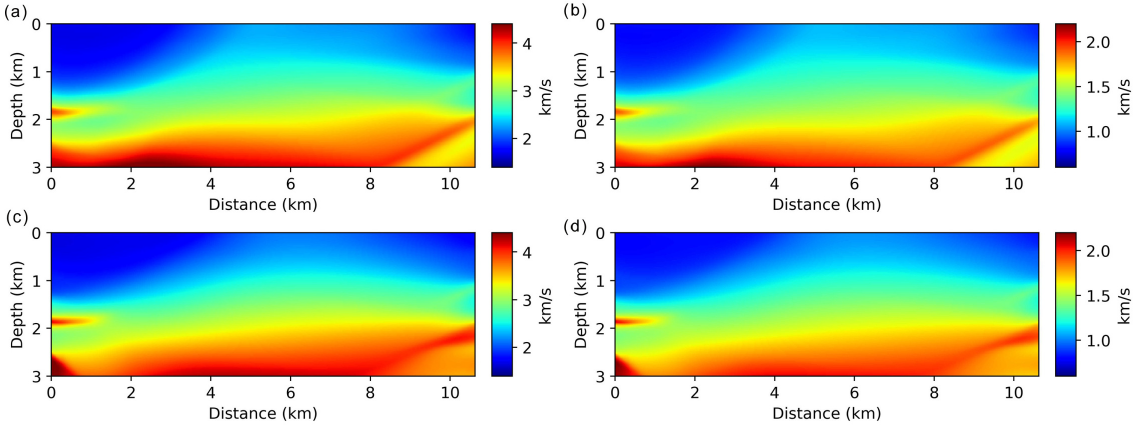} 
\caption{PINNPStomo-predicted (a) P-, (b) S-wave velocity models with the original factored form using true $v_{p0,s0}$; PINNPStomo-predicted (c) P-, (d) S-wave velocity models with the new factored form.}
\label{fig:mar_vp_vs_tomo_vsp_pred}
\end{center}   
\end{figure}

Then, we evaluate the efficacy of PINNPStomo in the P- and S-wave velocity inversion. When provided with the true $v_{p0,s0}$, the PINNPStomo-predicted P- and S-wave velocity models using the original factored eikonal equation are depicted in Figs. \ref{fig:mar_vp_vs_tomo_vsp_pred}a and \ref{fig:mar_vp_vs_tomo_vsp_pred}b, respectively. The background structures of the true models are successfully reconstructed without any prior information about the initial models. The new PINNPStomo method, which employs the new factored eikonal equation, demonstrates superior performance in recovering the background structures of the P- and S-wave inverted velocity models, particularly on the edges of the models where receivers are positioned, as illustrated in Figs. \ref{fig:mar_vp_vs_tomo_vsp_pred}c and \ref{fig:mar_vp_vs_tomo_vsp_pred}b, respectively.

\begin{figure}
\begin{center}
\includegraphics[width=1.0\textwidth]{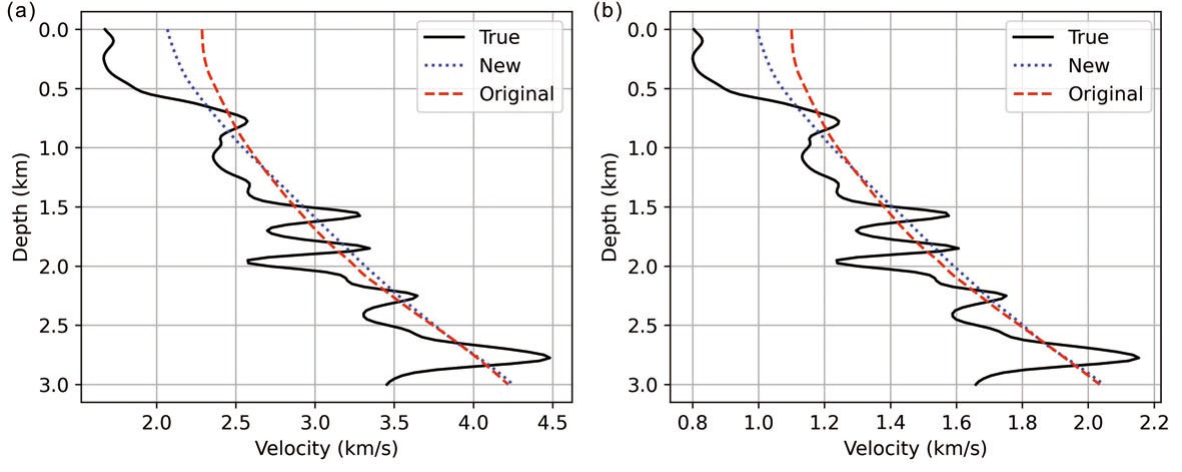} 
\caption{The vertical velocity profile comparisons for the (a) P- and (b) S-wave velocity models at the locations of $x=$ 5 km.}
\label{fig:mar_vp_vs_tomo_vsp_trace}
\end{center}   
\end{figure}

To further compare the performance of PINNPStomo between the original and new factored eikonal equations, we examine the vertical velocity profiles for the P- and S-wave velocity models at the location where $x=$ 5 km in Figs. \ref{fig:mar_vp_vs_tomo_vsp_trace}a and \ref{fig:mar_vp_vs_tomo_vsp_trace}b. It is evident that the new PINNPStomo strategy (blue dotted line) is more effective in recovering the low velocity in the shallow part compared to the original strategy (red dashed line). However, the inverted P- and S-wave velocity profiles from both strategies exhibit deviations from the true model (black solid line) in the shallow area. This discrepancy is attributed to the absence of receivers on the surface, which results in limited ray coverage of the shallow part.

\subsection{Crosswell data for 2D Overthrust model}

\begin{figure}
\begin{center}
\includegraphics[width=1.0\textwidth]{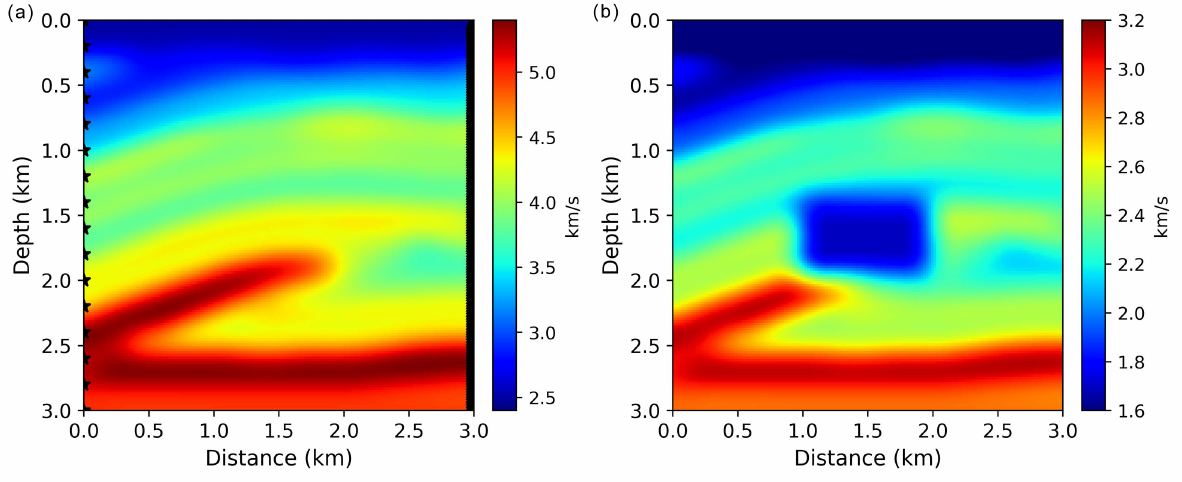} 
\caption{True (a) P- and (b) S-wave elastic Overthrust velocity models. The sources are placed in the left well indicated by the $\star$ sign; the receivers are distributed in the right well with a 25-m interval.}
\label{fig:over_true_crosswell}
\end{center}   
\end{figure}

Next, we apply the proposed method to the 2D Overthrust model with a crosswell acquisition geometry. The true modified P- and S-wave Overthrust velocity models are depicted in Figs. \ref{fig:over_true_crosswell}a and \ref{fig:over_true_crosswell}b, respectively. The sharp structures have been smoothed. To evaluate PINNPStomo's ability to mitigate the crosstalk issue, we introduce a low-velocity area in the middle of the S-wave velocity model. The model dimensions are $151 \times 151$, and its vertical and horizontal sampling is 20 m. We set 16 sources evenly distributed in the left well, indicated by the $\star$ sign in Fig. \ref{fig:over_true_crosswell}. Every grid point in the right well serves as a receiver.

\begin{figure}
\begin{center}
\includegraphics[width=1.0\textwidth]{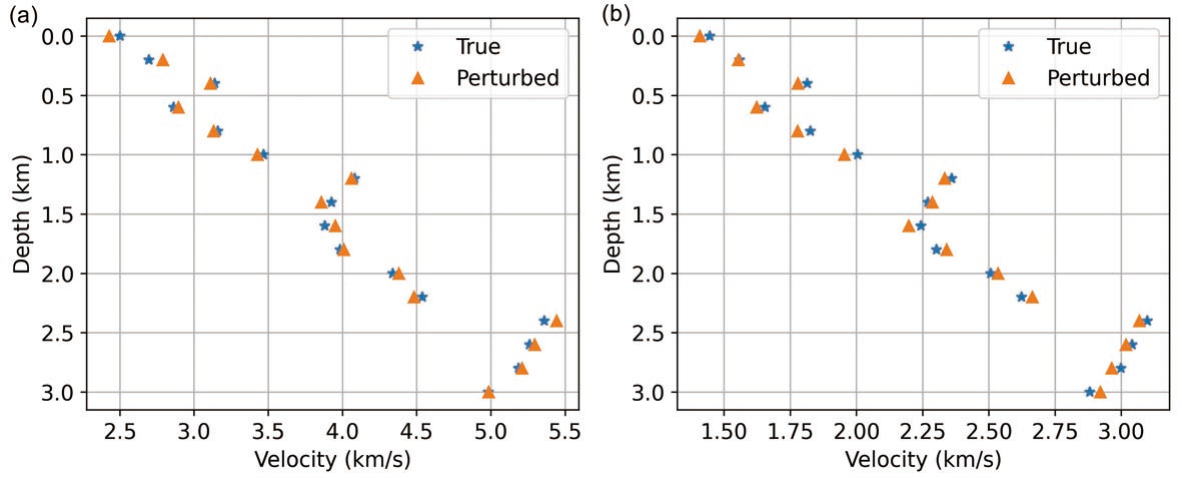} 
\caption{The true and perturbed (a) $v_{p0}$ and (b) $v_{s0}$ for the elastic Overthrust model.}
\label{fig:over_V_perturbed}
\end{center}   
\end{figure}

For this model, we use only 20,000 random sets of $\left \{ x,z,x_{z} \right \}$ as inputs to $\mathit{NN}1$. In this experiment, we train the networks with 500 epochs for three different PINNPStomo-based strategies. In the first one, assuming the P- and S-wave velocity values at the source locations ($v_{p0,s0}$) are obtained from the well log, we use the original factored eikonal (equation \ref{eqn:eq5}) as the physics constraint to train the networks. However, in reality, even the well-log velocity may not give the exact $v_{p0,s0}$ values for all the sources. Thus, for the second strategy, we add small random velocity perturbations to the true $v_{p0,s0}$ to calculate the background traveltimes $T_{p0,s0}$ used in the original factored eikonal equation. The maximum perturbed velocity values for $v_{p0}$ and $v_{s0}$ are 0.1 and 0.05, respectively. The true (blue star) and perturbed (orange triangle) $v_{p0,s0}$ are shown in Figs. \ref{fig:over_V_perturbed}a and \ref{fig:over_V_perturbed}b, respectively. In the last case, we use the proposed new factored eikonal (equation \ref{eqn:eq8}), which does not require the information of $v_{p0,s0}$, as the physics constraint in the loss function. 

\begin{figure}
\begin{center}
\includegraphics[width=0.75\textwidth]{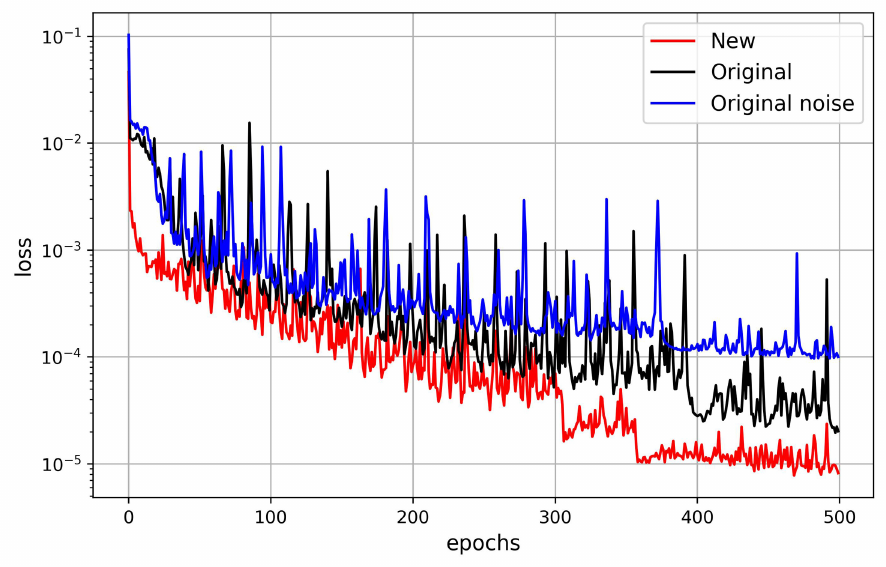} 
\caption{Training loss curves for the new factored eikonal equation and the original one with and without $v_{p0,s0}$ perturbation.}
\label{fig:over_loss_comp_vsp_random_noi}
\end{center}   
\end{figure}

The training loss curves for the three PINNPStomo strategies are displayed in Fig. \ref{fig:over_loss_comp_vsp_random_noi}. The new PINNPStomo method, which utilizes the new factored eikonal equation (red line), converges more rapidly to a lower loss compared to the original PINNPStomo that uses the true $v_{p0,s0}$ (black line). The training loss of the original PINNPStomo with perturbed $v_{p0,s0}$ does not converge as well as the strategy with the true $v_{p0,s0}$, as indicated by the blue line in Fig. \ref{fig:over_loss_comp_vsp_random_noi}.

\begin{figure}
\begin{center}
\includegraphics[width=1.0\textwidth]{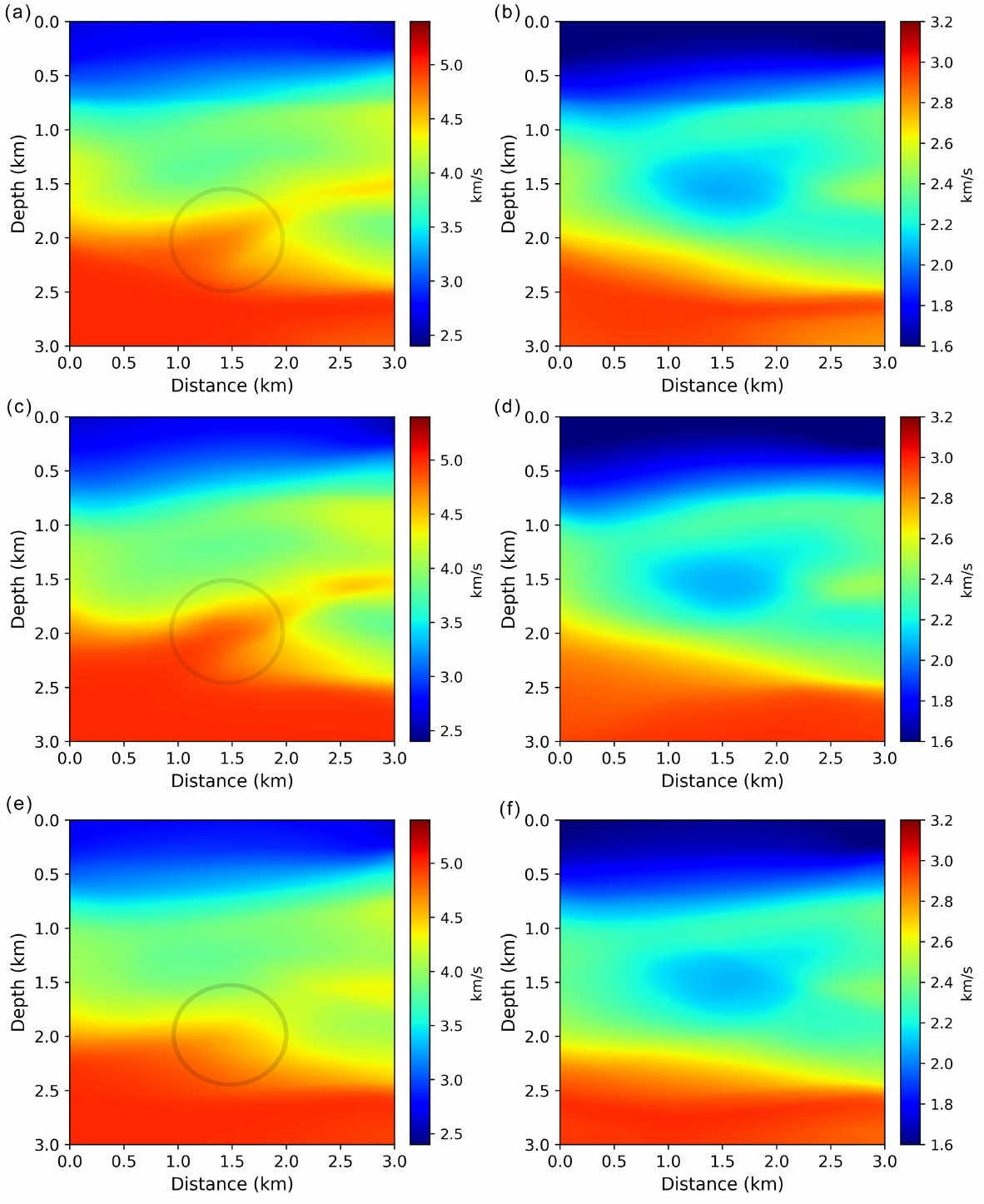} 
\caption{PINNPStomo-predicted (a) P-, (b) S-wave velocity models with the original factored form using true $v_{p0,s0}$; PINNPStomo-predicted (c) P-, (d) S-wave velocity models with the new factored form; PINNPStomo-predicted (e) P-, (f) S-wave velocity models with the original factored form using perturbed $v_{p0,s0}$. The circles in the inverted P-wave velocities column demonstrate the improved recovery of the high velocity layer by the new PINNPStomo (c) compared to the original ones (a), (e).}
\label{fig:over_vp_vs_tomo_crosswell_pred}
\end{center}   
\end{figure}

Figs. \ref{fig:over_vp_vs_tomo_crosswell_pred}a and \ref{fig:over_vp_vs_tomo_crosswell_pred}b display the PINNPStomo-predicted P- and S-wave velocity models by the original factored eikonal equation with true $v_{p0,s0}$. It is observed that the general structure of the true model is recovered, and the low-velocity area in the middle of the S-wave model is successfully captured. PINNPStomo with the new factored eikonal equation also manages to recover the structure of the true P- and S-wave velocity models, with a noticeable enhancement in the high-velocity layer (circled area), as displayed in Figs. \ref{fig:over_vp_vs_tomo_crosswell_pred}c and \ref{fig:over_vp_vs_tomo_crosswell_pred}d. Despite adding very small random perturbations to the true $v_{p0,s0}$, the quality of the PINNPStomo-predicted P-wave velocity obtained using the original factored eikonal equation with perturbed $v_{p0,s0}$ values decreases significantly in the circled area, as demonstrated in Figs. \ref{fig:over_vp_vs_tomo_crosswell_pred}e. This highlights the significance of having exact $v_{p0,s0}$ values in the original PINNPStomo. However, acquiring such precise values is highly challenging in realistic seismic surveys.

\begin{figure}
\begin{center}
\includegraphics[width=1.0\textwidth]{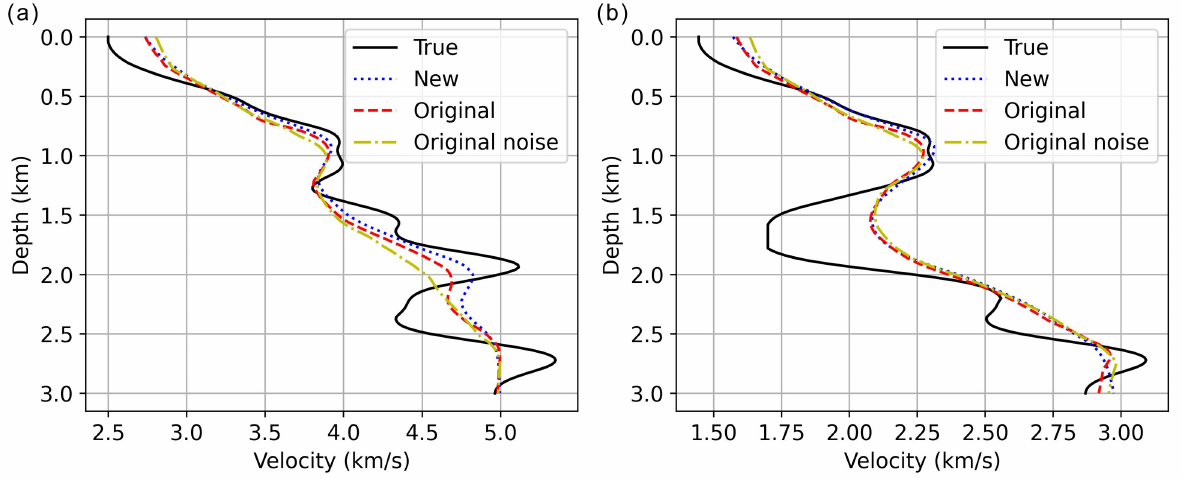} 
\caption{The vertical velocity profile comparisons for the (a) P- and (b) S-wave velocity models at the locations of $x=$ 1.5 km.}
\label{fig:over_V_trace_tomo75}
\end{center}   
\end{figure}

We compare the vertical P- and S-wave velocity profiles generated by the new and original PINNPStomo at the locations where $x=$ 1.5 km, as illustrated in Figs. \ref{fig:over_V_trace_tomo75}a and \ref{fig:over_V_trace_tomo75}b. In the P-wave velocity profile, it is evident that the new PINNPStomo (blue dotted line) provides a superior reconstruction of the high-velocity layer at a depth of 2 km compared to the original PINNPStomo (red dashed line). Similarly, in the S-wave velocity profile, the PINNPStomo-predicted velocity using the new eikonal equation (blue dotted line) more closely matches the true velocity at a depth of 1 km than the original PINNPStomo (red dashed line). If the perturbed $v_{p0,s0}$ is used for the original PINNPStomo, the original PINNPStomo fails to capture the high-velocity layer for the P-wave velocity (green dashed-dotted line).

\subsection{Passive seismic data for 3D Overthrust model}

\begin{figure}
\begin{center}
\includegraphics[width=1.0\textwidth]{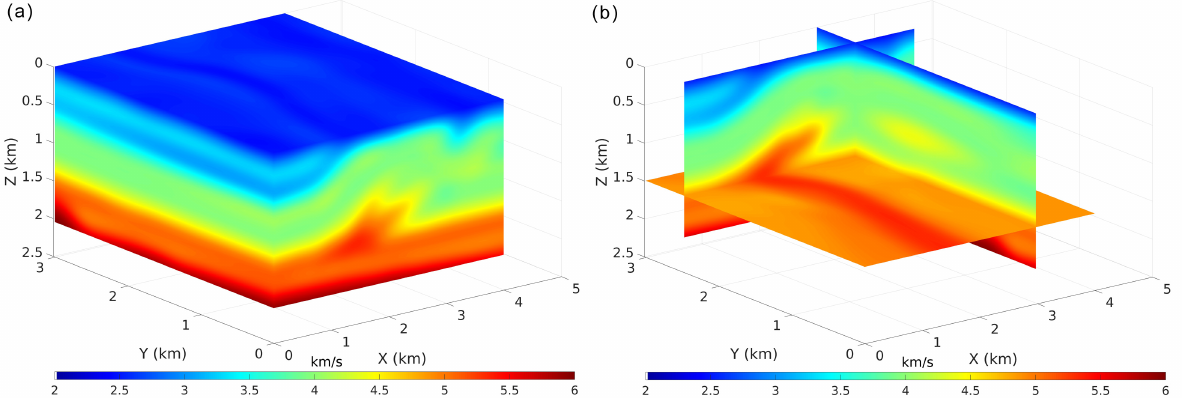} 
\caption{True 3D P-wave elastic Overthrust velocity models and (b) its three internal slices.}
\label{fig:over3d_vp_true}
\end{center}   
\end{figure}

\begin{figure}
\begin{center}
\includegraphics[width=1.0\textwidth]{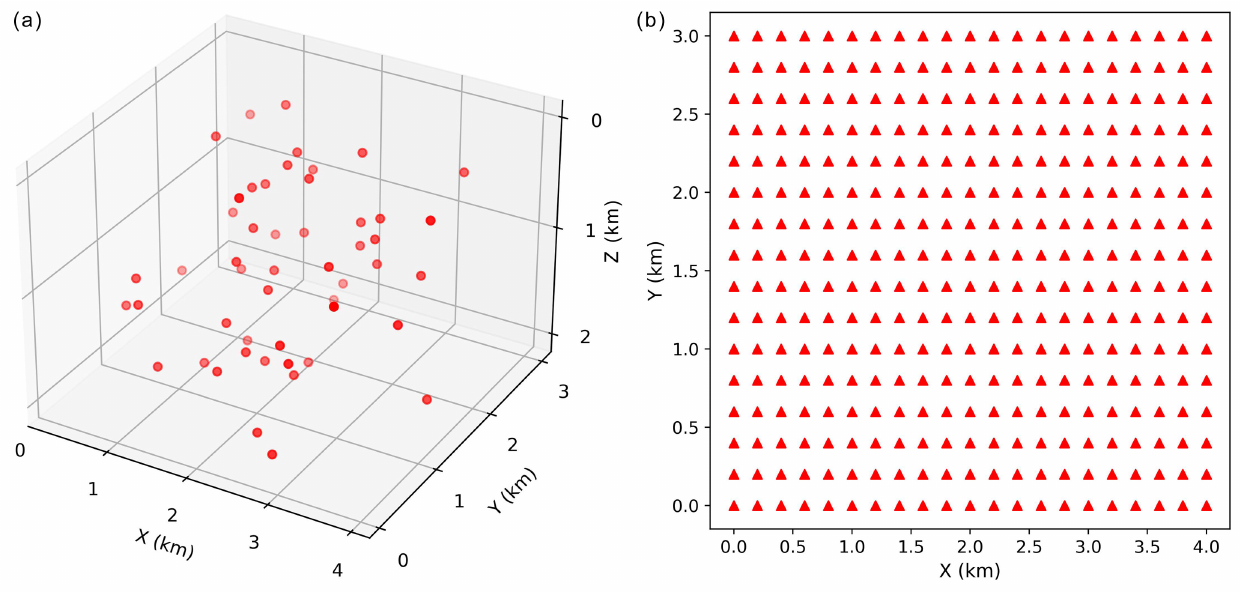} 
\caption{The locations of the (a) passive sources and (b) surface receivers.}
\label{fig:over3d_sources}
\end{center}   
\end{figure}

Finally, we test the proposed method on a modified elastic 3D Overthrust model.  The P-wave velocity and three internal slices of the model are displayed in Figs. \ref{fig:over3d_vp_true}a and \ref{fig:over3d_vp_true}b, respectively. The S-wave velocity is derived by dividing the P-wave velocity by a constant value of 1.731.The model dimensions are $161 \times 121 \times 83$ with a spatial sampling of 25 meters. We assume there are 50 passive seismic events randomly distributed in the subsurface, which have been located by passive source estimation methods. Their locations are indicated in Fig. \ref{fig:over3d_sources}a. For this 3D seismic survey, we establish a relatively sparse data acquisition geometry on the surface, consisting of a total of 336 receivers, distributed as shown in Fig. \ref{fig:over3d_sources}b. This seismic geometry setup is comparable to P- and S-wave traveltime tomography at regional and global scales.

\begin{figure}
\begin{center}
\includegraphics[width=0.75\textwidth]{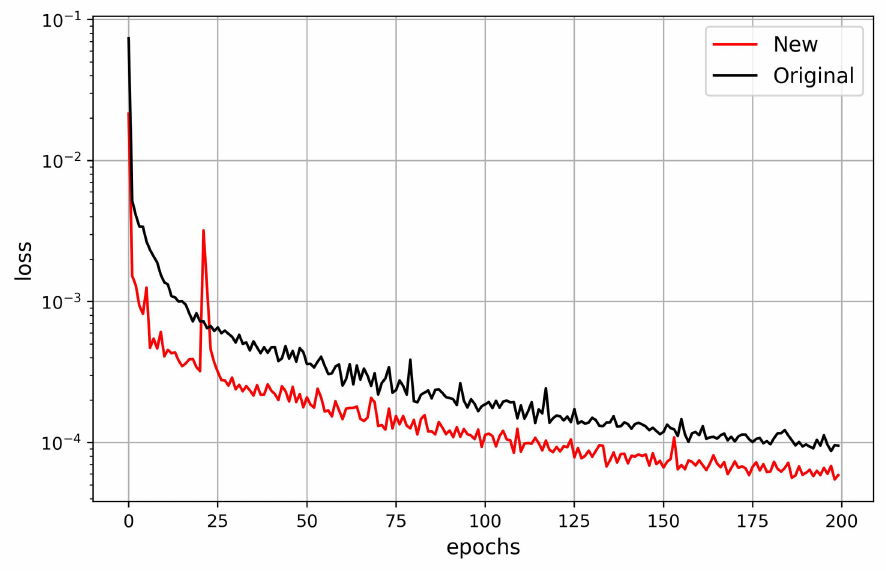} 
\caption{Training loss curves for the new factored eikonal equation and the original one.}
\label{fig:over3d_loss}
\end{center}   
\end{figure}

For the inversion of this 3D P- and S-wave velocity models, we input 500,000 sets of $\left \{ x, y, z, x_{s}, y_{s}, z_{s} \right \}$consisting of random values within the model domain's range into $\mathit{NN}1$, and we input the corresponding $\left \{ x, y, z \right \}$ to $\mathit{NN}2$. We train PINNPStomo for only 200 epochs. The training loss curves for PINNPStomo with the new factored eikonal equation and the original version are presented in Fig. \ref{fig:over3d_loss}. We reach the same conclusion that PINNPStomo converges more effectively when using the new factored eikonal equation (red line) compared to the original equation (black line).

\begin{figure}
\begin{center}
\includegraphics[width=1.0\textwidth]{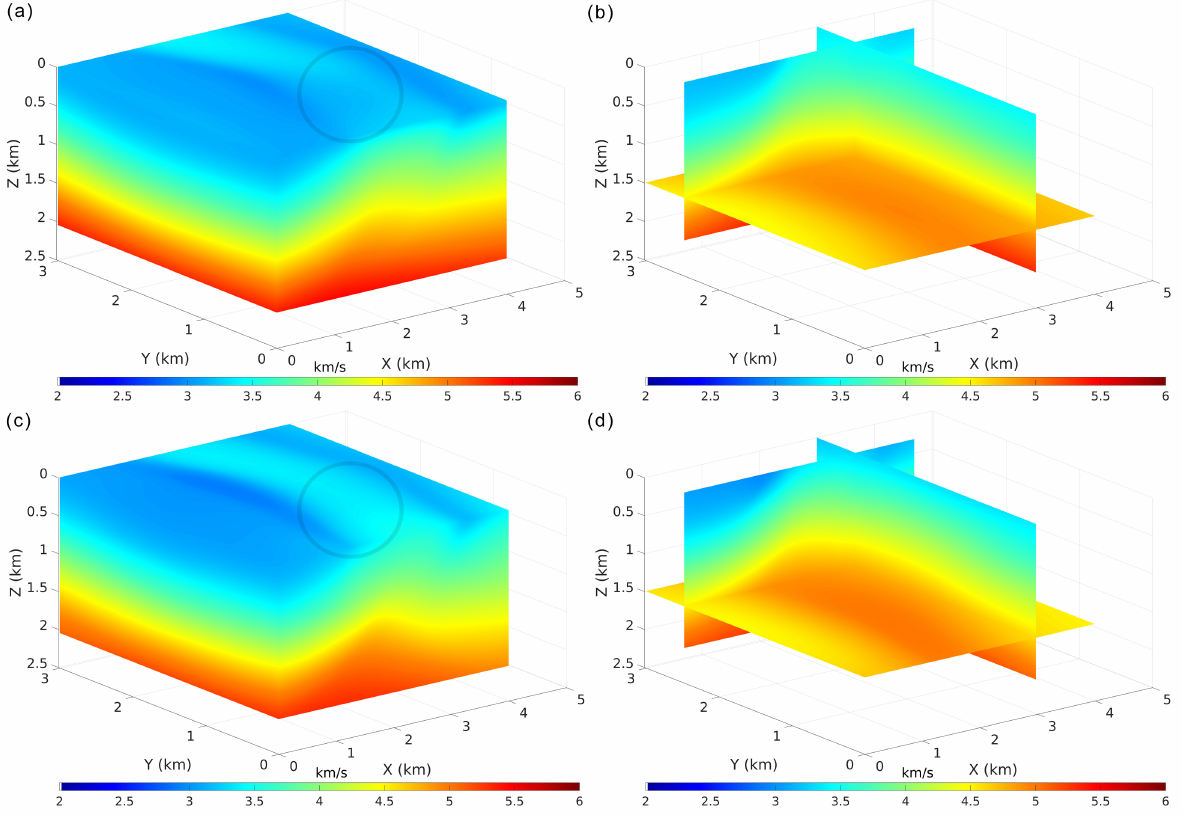} 
\caption{PINNPStomo-predicted P-wave velocity models with the (a),(b) original and (c),(d) new factored eikonal equation.}
\label{fig:over3d_vp_pred}
\end{center}   
\end{figure}

Figs. \ref{fig:over3d_vp_pred}a and \ref{fig:over3d_vp_pred}b present the PINNPStomo-predicted P-wave velocity models using the original factored eikonal equation. The general background structures of the true model are well reconstructed. In comparison, the proposed new PINNPStomo method further enhances the accuracy of the inverted P-wave velocity model, as indicated by the circled area in Fig. \ref{fig:over3d_vp_pred}c. Since the S-wave velocity shares the same structure as the P-wave velocity, we only display the P-wave velocity inversion results in this example.

\begin{figure}
\begin{center}
\includegraphics[width=1.0\textwidth]{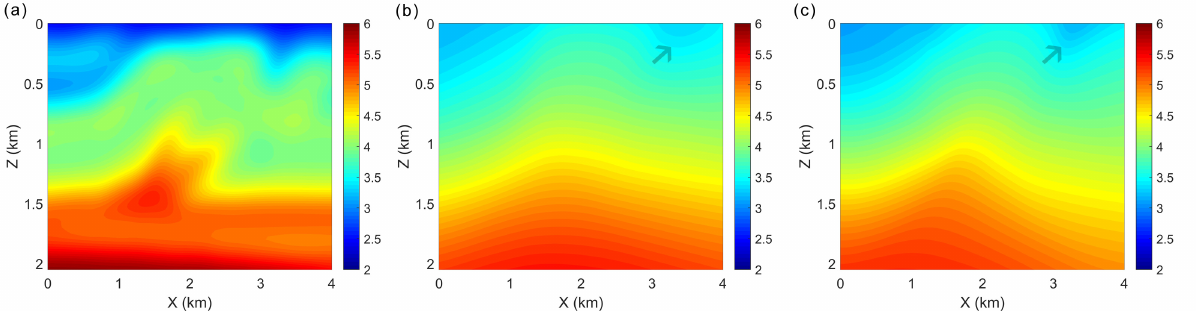} 
\caption{The vertical velocity slices at $y=$ 0 km of the (a) true velocity model, PINNPStomo-predicted P-wave velocity with the (b) original and (c) new factored eikonal equation.}
\label{fig:over3d_vp_xz}
\end{center}   
\end{figure}

\begin{figure}
\begin{center}
\includegraphics[width=1.0\textwidth]{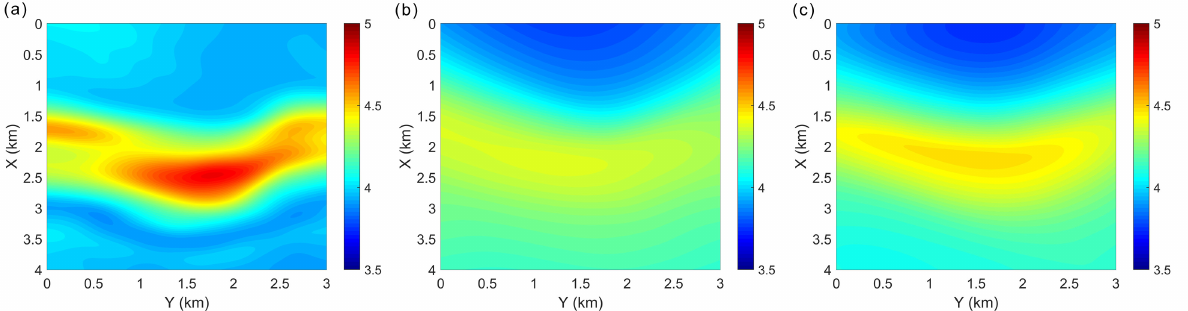} 
\caption{The horizontal velocity slices at the depth of $z=$ 1 km of the (a) true velocity model, PINNPStomo-predicted P-wave velocity with the (b) original and (c) new factored eikonal equation.}
\label{fig:over3d_vp_xy}
\end{center}   
\end{figure}

To conduct a detailed comparison between the inversion results from the new and original PINNPStomo, we extracted one vertical slice at $y=$ 0 km and one horizontal slice at $z=$ 1 km. Figs. \ref{fig:over3d_vp_xz}a,  \ref{fig:over3d_vp_xz}b, and \ref{fig:over3d_vp_xz}c show the true, original PINNPStomo-predicted, and new PINNPStomo-predicted vertical velocity slices, respectively. We can observe that the new PINNPStomo provides a better reconstruction of the low-velocity structure in the shallow part, as indicated by the arrows in Figs. \ref{fig:over3d_vp_xz}b and \ref{fig:over3d_vp_xz}c.  Figs. \ref{fig:over3d_vp_xy}a,  \ref{fig:over3d_vp_xy}b, and \ref{fig:over3d_vp_xy}c display the true, original PINNPStomo-predicted, and new PINNPStomo-predicted horizontal velocity slices, respectively. It is evident that the new PINNPStomo successfully captures the high-velocity layer in the middle.

\section*{Discussion}

Simultaneous inversion of P- and S-wave velocity models is crucial for seismic exploration and imaging the interior of the Earth. From the perspective of exploration seismology, the shorter wavelength of S waves can enhance the resolution of subsurface images compared to P-wave inversion alone, which is vital for detailed structural interpretation. Furthermore, P- and S-wave velocities can be used to ascertain the elastic properties of rocks, which is essential for identifying oil and gas reservoirs. However, picking the first-arrival traveltimes of S waves can be challenging due to the strong energy of surface waves, a common issue in land seismic data. In VSP and crosswell data, surface waves are less likely to be recorded, allowing for the picking of first-arrivals for both P- and S-waves. The 2D examples presented in this paper demonstrate that the proposed PINNPStomo method can successfully perform P- and S-wave velocity inversion without the crosstalk issue.

From the regional or global seismology perspective, simultaneous inversion of P- and S-wave velocities aids in interpreting the composition and state of the Earth's interior, encompassing the crust, mantle, and core. In seismic tomography, the first arrivals of P- and S-waves from earthquakes within the Earth can be identified. The 3D model example discussed in this paper illustrates that the proposed method holds significant potential for P- and S-wave tomography in addressing regional and global seismology problems.

The original PINNtomo method employs the multiplicative factored eikonal equation, which relies on prior information about the velocity values at the source locations. Given the nature of crosswell and passive seismic surveys, obtaining exact velocity values in the subsurface is challenging. To eliminate this dependency on source velocity, we abandon the usage of the background traveltime $T_{0}$ and instead compute the traveltime $T$ by multiplying the seismic wave travelling distance $\left | \mathbf{x}-\mathbf{x_{s}} \right |$ with a factor $\gamma$. $\gamma$ can be considered as the average slowness between the source location $\mathbf{x_{s}}$ and the grid point $\mathbf{x}$ with the unit of $s/km$. The range of $\gamma$ is $\left [ \frac{1}{v_{max}}, \frac{1}{v_{min}} \right ]$ for all sources. In contrast, in the original multiplicative factored eikonal equation, the range of $\tau$ is $\left [ \frac{v_{0}}{v_{max}}, \frac{v_{0}}{v_{min}} \right ]$, which varies with $v_{0}$. Consequently, it is easier for the neural networks to search for $\gamma$ within a consistent range. This explains the faster convergence of PINNPStomo with the new factored eikonal equation. In principle, the new PINNPStomo is equivalent to the original one when $v_{0}=1$ for all sources. However, the new PINNPStomo is more physically interpretable and simpler to implement since it does not require the calculation of $T_{0}$. 

The proposed PINNPStomo framework has two main advantages. The first one is its independency from initial models for reconstructing P- and S-wave velocities. In PINNPStomo, velocities are reparameterized by the trainable parameters of neural networks. The inversion process begins with the random initialization of neural networks rather than initial velocities. The second advantage is its efficiency in simultaneously inverting both P- and S-wave velocities with a single training session. We use one NVIDIA RTX A6000 to train the networks. For the 3D model, it takes only 22.1 minutes to complete the training and obtain the 3D inverted P- and S-wave velocity models. Currently, we do not have field datasets available for testing. In the future, we plan to verify the effectiveness and efficiency of PINNPStomo on available large-scale field datasets.

\section*{Conclusion}

We have developed an innovative method for simultaneous P- and S-wave traveltime tomography using physics-informed neural networks, known as PINNPStomo. PINNPStomo utilizes two separate neural networks to represent the P-, S-wave traveltimes and velocities. By minimizing the physics loss constrained by the eikonal equation and the data loss from the traveltime misfit, we can optimize these two networks and obtain the predicted P- and S-wave velocities simultaneously. To reduce the dependency on velocities at the source locations in the original multiplicative factored eikonal equation, we introduced a new factored eikonal equation that eliminates the need to calculate the background traveltime and incorporated it into PINNPStomo. This new version of PINNPStomo further enhances the convergence performance during training, leading to higher accuracy in reconstructing P- and S-wave velocities. We have validated its effectiveness and superiority using 2D and 3D benchmark velocity models under various seismic acquisition geometry setups.

\section*{Data and code availability}

The related codes for researchers to evaluate the proposed method have been uploaded to https://github.com/songc0a/PINNPStomo/.

\section{Acknowledgement}

We thank Jilin University for its support. 

\bibliographystyle{unsrt}  
\bibliography{refs/pinn}

\end{document}